\documentclass{elsart}
\usepackage{natbib}
\usepackage{epsfig}
\usepackage{unites2e}
\begin{document}
\runauthor{LeBohec}
\begin{frontmatter}
\title{Minimal Stereoscopic Analysis for Imaging Atmospheric Cherenkov 
Telescope Arrays}

\author[MCSD]{S. LeBohec,}
\author[GRIN]{C. Duke,}
\author[PJ]{and P. Jordan}
\author[PJNOW]{}
\address[MCSD]{Department of Physics, University of Utah\\
               Salt-Lake-City, UT, 84112-0830, USA}
\address[GRIN]{Department of Physics, Grinnell College, \\
               Grinnell, IA, 50112-1690}
\address[PJ]{Department of Physics and Astronomy, Iowa State University\\
  Ames, IA, 50010-3160}
 \address[PJNOW]{Now in the Department of Financial Engineering, University 
of Michigan,\\ Ann Arbor, MI 48109-2092}
    
\begin{abstract}
The trajectory of a primary $\gamma$-ray detected with an array of at least 
four atmospheric Cherenkov imaging telescopes can be reconstructed from the 
shower image centroid positions and geometrical considerations independent 
of the primary energy. Using only the image centroid positions some cosmic-ray 
discrimination is also possible. This minimal approach opens the possibility 
of pushing the analysis threshold to lower values, close to the hardware 
threshold. 

\vspace{1pc}
\end{abstract}
\end{frontmatter}

\section{Introduction}
 The rapid development of ground-based high-energy astronomy during
 the last ten years \cite{weekes03} is closely  associated with the
 success of the Atmospheric Cherenkov Imaging technique \cite{hillas96}
 developed with the Whipple Observatory 10m telescope
 \cite{weekes89} and elsewhere. Cosmic rays and $\gamma$-rays entering the 
 atmosphere generate cascades of particles   
radiating Cherenkov light. At ground level, the Cherenkov light front 
extends over more than $\rm \sim 130m$ from the shower axis. Thus, a 
single 
imaging telescope achieves an effective collection area of more than 
$\rm 5\times10^4m^2$. The shape of the image can be used to
 preferentially select $\gamma$-ray events over the much more frequent 
cosmic-ray events. Furthermore, the elongated image of a $\gamma$-ray shower 
points back to the source on the sky, providing information on its arrival 
direction.

Most of the present generation experiments (CANGAROO, HESS, MAGIC and VERITAS) 
consist of arrays of imaging telescopes. Following earlier pioneering
studies \cite{krennrich98, chadwick96}, the power of the stereoscopic 
technique \cite{grindlay75}, was plainly illustrated with the HEGRA  
experiment \cite{kohnle96, pulhofer03} and more recently with the HESS
experiment \cite{hinton04} which achieved unprecedented levels of $\gamma$-ray 
discrimination. 

In these arrays, each telescope records an image of the same shower. For a 
$\gamma$-ray primary, the major axis of each image points
back to the source so that superposing the images produces the
source location on the sky. Furthermore, the position of the image around 
the source corresponds to the position of the shower around the line from the 
telescope to the source. Combining this information from two or more telescopes
provides the position and orientation of the shower axis. In other words, 
each telescope can
be associated with a plane that contains the shower axis and the 
telescope. As the shower axis is the line of intersection of the planes
associated with each of the array telescopes, only two telescopes are required 
to extract this axis. Since accurate reconstruction is difficult or
impossible when the impact point is closely aligned with the two
telescopes, a minimum of three telescopes is preferable.

The stereoscopic technique described above requires that each image 
contains a sufficient number of pixels for the major axis to be 
clearly identified. 
However, this is generally not the case for events close to the detection 
threshold and for which some or even all of the images can consist of only a 
few pixels significantly above noise fluctuations. Only the image positions 
are then available for a reconstructive analysis. In this paper, we study the
possibility of reconstructing $\gamma$-ray events using only the shower image 
centroid position from each telescope. Such a minimalist analysis might be 
particularly important for the lowest energy events. The next section 
describes the method and in the following section, we use Monte-Carlo 
simulations to present an illustration of the method. 
Finally, we discuss approaches for discriminating against cosmic-rays when 
the image positions are the only available significant information.

\section{Stereoscopy with minimal imaging}
 \begin{figure}[geom]
\epsfig{file=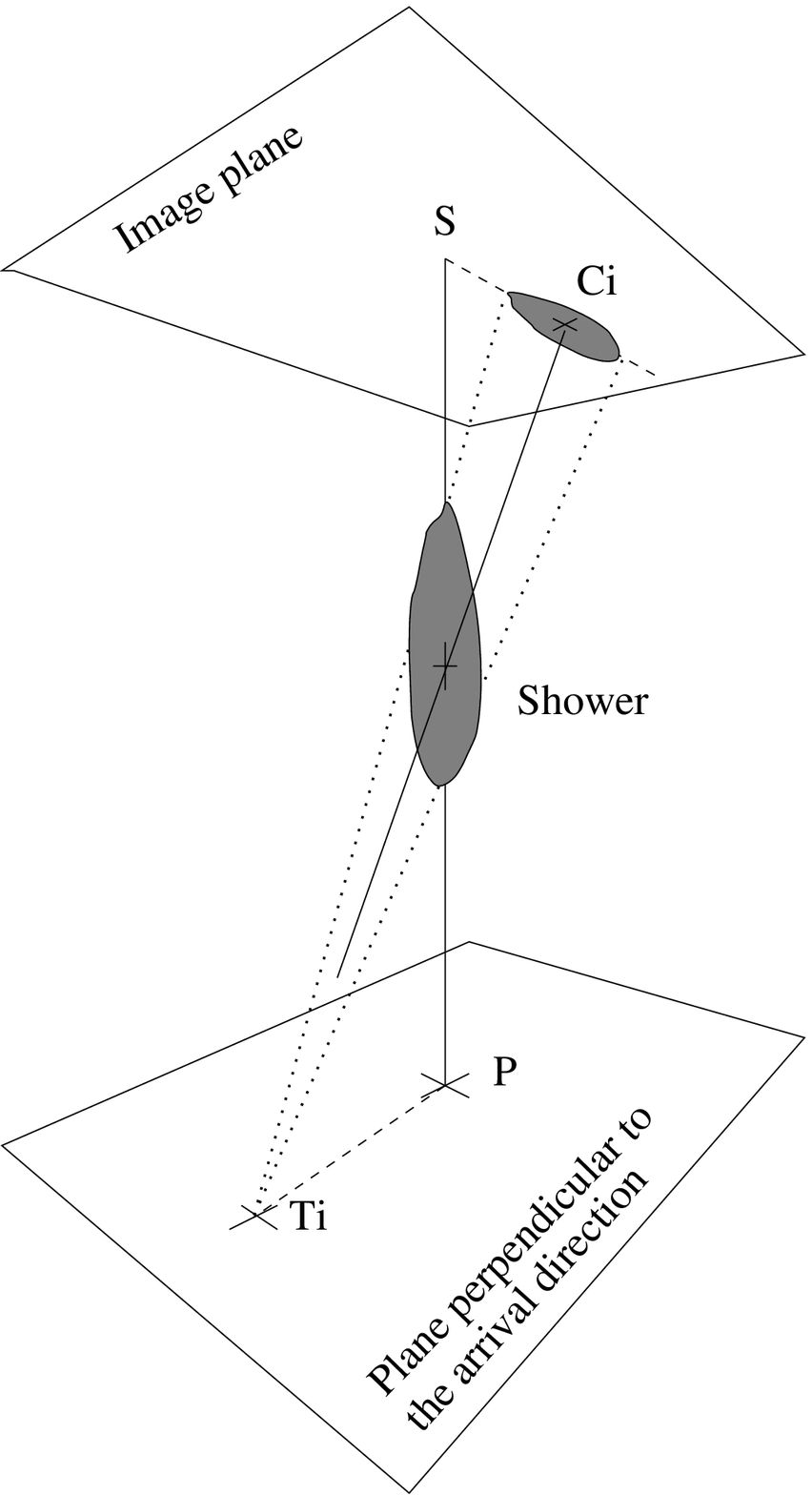,width=2.80in}
\vspace{10pt}
\caption{The telescope $\rm T_i$ and the shower axis define a plane that 
contains the source and the image centroid.}  
\label{geom}
\end{figure}
Consider a plane perpendicular to the shower axis. Let $\rm P$ be the
intersection point of the shower axis in this plane. Project the location of
each telescope along the direction of the shower axis to the plane; label
these points $\rm T_i$. Locate the source, $\rm S$, and  
the image centroid, $\rm C_i$, on the sky. We assume the image centroid to lie 
on the shower axis projected on the sky so that $\rm T_i$, $\rm P$, $\rm C_i$ 
and S are coplanar as shown on figure \ref{geom}. As a result the
lines, $\rm (SC_i)$ and  
$\rm (PT_i)$, are parallel. Expressing this result using convenient
coordinate systems on the sky and in the plane perpendicular to the source
direction yields:  

\begin{equation}
\rm (x_P-x_{T_i})\cdot(y_S-y_{C_i})-(y_P-y_{T_i})\cdot(x_S-x_{C_i})=0
\label{equat}
\end{equation}

where i ranges from 1 though the number of telescopes, N. Since there
are four unknown quantities, N must generally be at least equal 
to four for the system to have a discrete number of solutions.
A minimum of four telescopes is necessary for reconstructing the shower axis
from the image centroids. In fact, the minimal telescope number could in 
principle be reduced to three or even two through the geometrical connection 
between the distance $\rm (SC_i)$ and the impact parameter to telescope 
$\rm i$. However, this relationship, by relying on  the physics as well as the 
geometry of the atmospheric shower, depends on the
primary energy and is more susceptible to fluctuations than our assumption 
that the image centroid lies on the shower
axis image. We prefer not to use such a complicated relationship at this 
stage. 

Because of the projection of the telescope locations onto the
plane perpendicular to the primary direction, $\rm x_P$ and $\rm y_P$ 
depend on the primary arrival direction. However, 
we note that the field of view of atmospheric Cherenkov imaging telescopes 
remains relatively small, less than 2 or 3 degrees in radius. Thus, to a good 
approximation, we can perform the projection along the tracked 
direction instead of the arrival direction. For a typical array with 
distances between telescopes of the order of 100m, this approximation will 
affect the relative positions of the telescopes in the projection plane by 
much less than the physical size of the telescopes.
  
We can now directly use, for example, the first two equations in 
\ref{equat} to 
express $\rm x_P$ and $\rm y_P$ as functions of $\rm x_S$ and $\rm y_S$. These 
expressions can then be substituted into the remaining equations. 
For more than four telescopes, the system is over-constrained. In the case 
of a four-telescope array, the system consists of two coupled quadratic 
equations in  $\rm x_S$ and $\rm y_S$ with no apparent analytical solution. 
We thus use a numerical approach to find the source and impact point by 
minimizing the function: 
\begin{equation}
\rm f(x_S,y_S)=\sum_{i=3}^{i=N} ((x_P-x_{T_i})\cdot(y_S-y_{C_i})-(y_P-y_{T_i})\cdot(x_S-x_{C_i}))^2
\end{equation}.

\begin{figure}[stereoprincip]
\epsfig{file=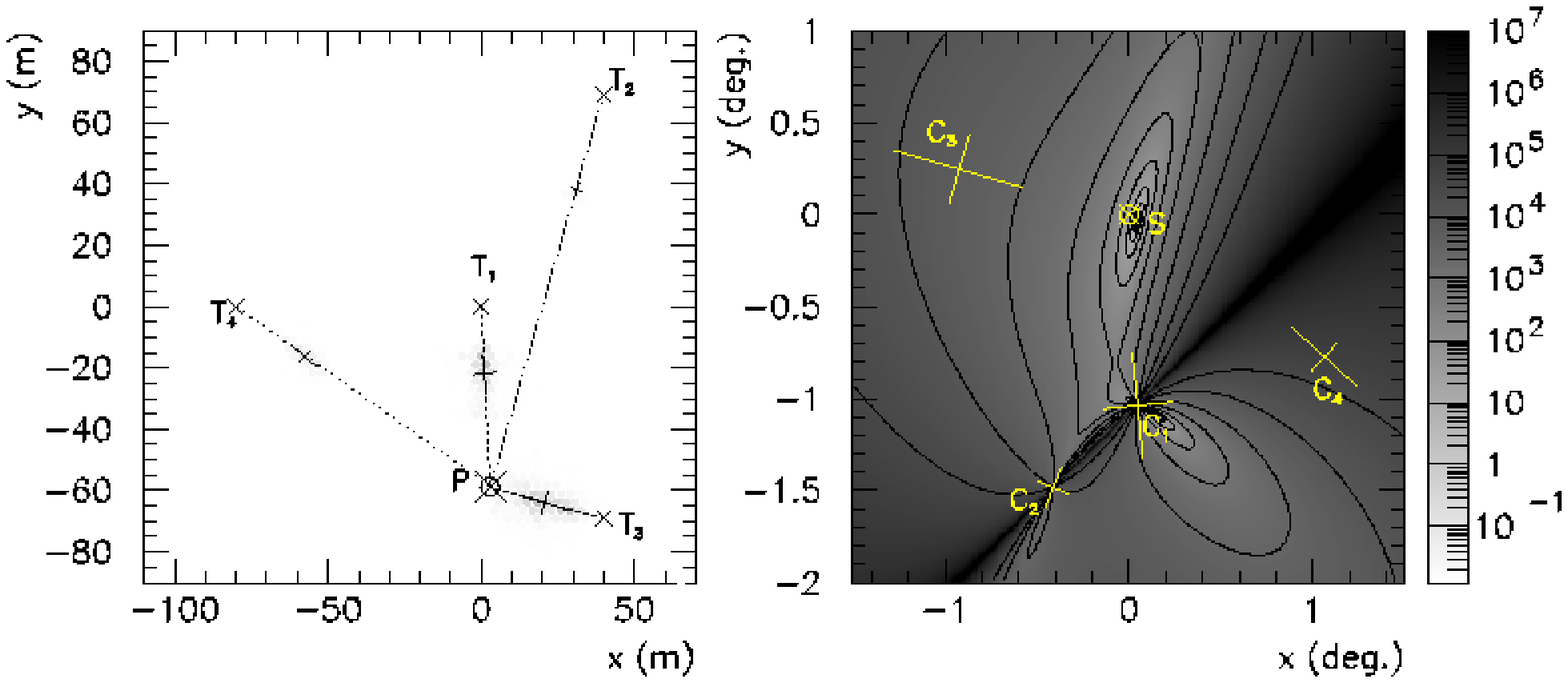,width=5.5in}
\vspace{10pt}
\caption{On the left figure, in a plane perpendicular to the simulated 
$\gamma$-ray arrival direction, $\rm T_1$, $\rm T_2$, $\rm T_3$ 
and $\rm T_4$ locate the telescopes. $\rm P$ is the point where the 
primary would have arrived. Each Cherenkov image is aligned with P and the 
corresponding telescope. On the right side, in the image plane, $\rm C_1$, 
$\rm C_2$, $\rm C_3$ and $\rm C_4$ are the image centroids and $\rm S$
is the actual source position. The function $\rm f(x_S,y_S)$ is displayed 
with a logarithmic grey scale and contour lines.}
\label{stereoprincip}
\end{figure}

Since the equations for telescopes 1 and 2 are used to express $\rm x_P$ and 
$\rm y_P$, the sum extends from $\rm i=3$ up to $\rm i=N$, the total 
number of telescopes. This function reaches a minimum at the source 
position. In practice, by further manipulating the equations, the 
minimization becomes one-dimensional. For an array of more 
than four telescopes ($\rm N>4$), two approaches are possible: a full 
four-dimensional minimization or one-dimensional minimizations for all groups 
of three telescopes. In the latter case, the solution will depend on the
specific order of equation choices in reducing the problem to a
one-dimensional optimization. 
The  ${\rm N \choose 3}$ solutions have to be combined into a single solution. 
Our examples  only use this one-dimensional minimization with a 
four-telescope-array. 
Figure \ref{stereoprincip}  illustrates  
the geometry and the $\rm f(x_S,y_S)$ function for a simulated vertical 
500~GeV shower and a four-telescope array ($\rm N=4$) arranged in a centered 
equilateral triangle. Because of the high primary energy, the major axes of 
the images precisely point toward the source position and impact
point.
 
The function $\rm f(x_S,y_S)$ reaches a minimum near $\rm S$ because the 
lines connecting $\rm S$ to the various image centroids are approximately 
parallel to the lines connecting $P$ to the various telescopes. In the 
following section, we present the capabilities of this method for this array.

\section{Example Application}
\subsection{Reconstruction capabilities}
 \begin{figure}[perform]
\epsfig{file=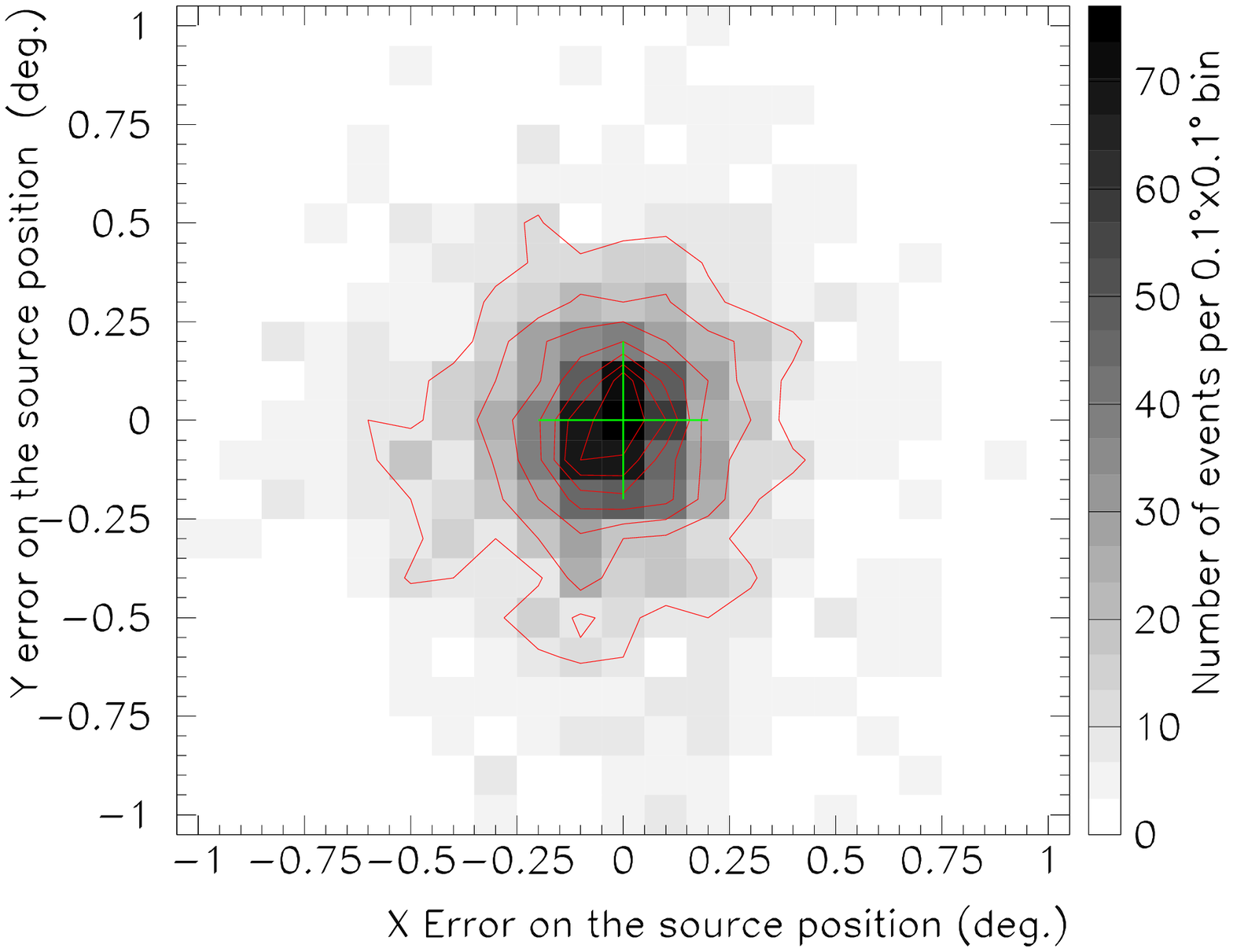,width=2.80in}
\epsfig{file=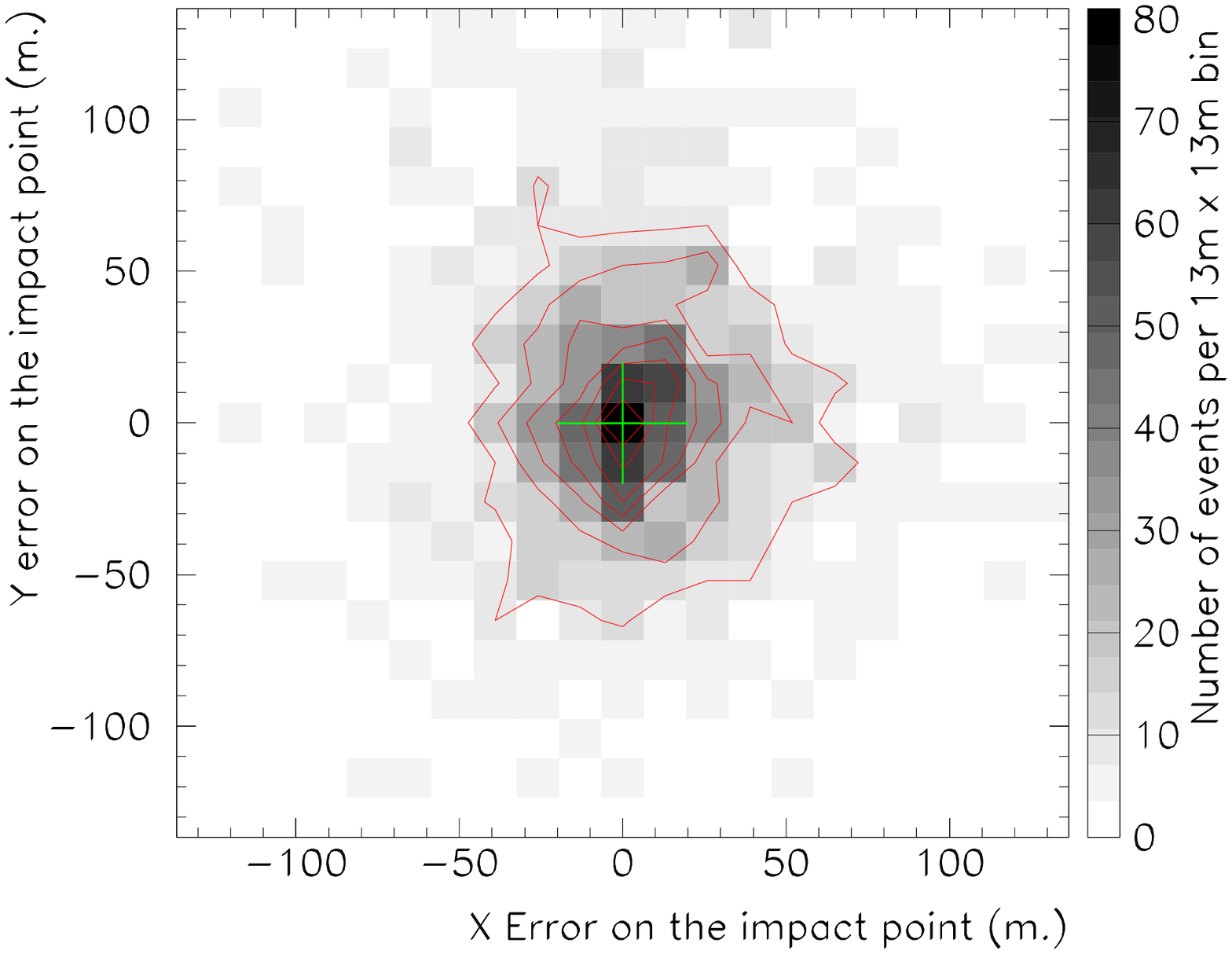,width=2.80in}
\vspace{10pt}
\caption{Distributions of the errors in source position (left) and
impact point (right) reconstructions for 100~GeV $\gamma$-rays and a
VERITAS like array.}  
\label{perform}
\end{figure}

In the case of a four-telescope array, the function $\rm f(x_S,y_S)$ reaches 
zero in two points. On figure \ref{stereoprincip}, one solution is close to 
the actual source position; the second solution is near $\rm C_1$. 
A fifth telescope could in principle select the proper 
solution, but we can also include additional, as yet unused,
information. For example, each image centroid should lie on the 
the impact position side of the source point 
as indicated by the spatial position of the impact point with
respect to the corresponding telescope. 
Unfortunately, this criterion does not always select only one of the two the
solutions. As previously discussed, we have not incorporated the distance
$\rm (SC_i)$ from the source to the image centroid. For
$\gamma$-rays, this distance is approximately linearly  
correlated with the impact parameter $\rm (T_iP)$. 
The slope of the relationship depends on the elevation and primary energy. 
Hence, we can require that the ratio, 
$\rm r_i= (SC_i)/(T_iP)$ have approximately the same value for
each telescope. In practice, when the first criterion does not select a
unique solution, we compute the standard deviation of the
four distance ratios for both solutions and retain the solution with
the smallest standard deviation. 

To test and illustrate this method, we used the GrISU simulation 
package \cite{grisu}  to simulate vertical 100-GeV $\gamma$ rays with  
a VERITAS-like array of four 12m-telescopes.  Each shower image is 
polluted by a typical level of night sky background. The individual 
telescope trigger 
conditions demanded that three channels coincidently exceed a threshold 
corresponding to four photoelectrons. All four telescopes must trigger
to record an event. These are detailed Monte-Carlo simulations
generated for a generic ACT array; hence, they do not include the level
of specificity required to characterize a particular instrument.

Figure \ref{perform} shows the error distribution
of the source position and 
impact-point reconstruction. The standard deviation from 
the actual source position is  $\rm 0.2^o$ and the standard deviation from the
actual impact point is  $\rm 20~m$. 
This analysis only uses the image 
centroids. We realize that an analysis using the distribution of light 
in the image, when the image is sufficiently bright, might give more
accurate results. However, it is important to note that no threshold
nor cosmic-ray discrimination  
criteria have been applied yet. 

\subsection{Discrimination possibilities}
 \begin{figure}[discri]
\epsfig{file=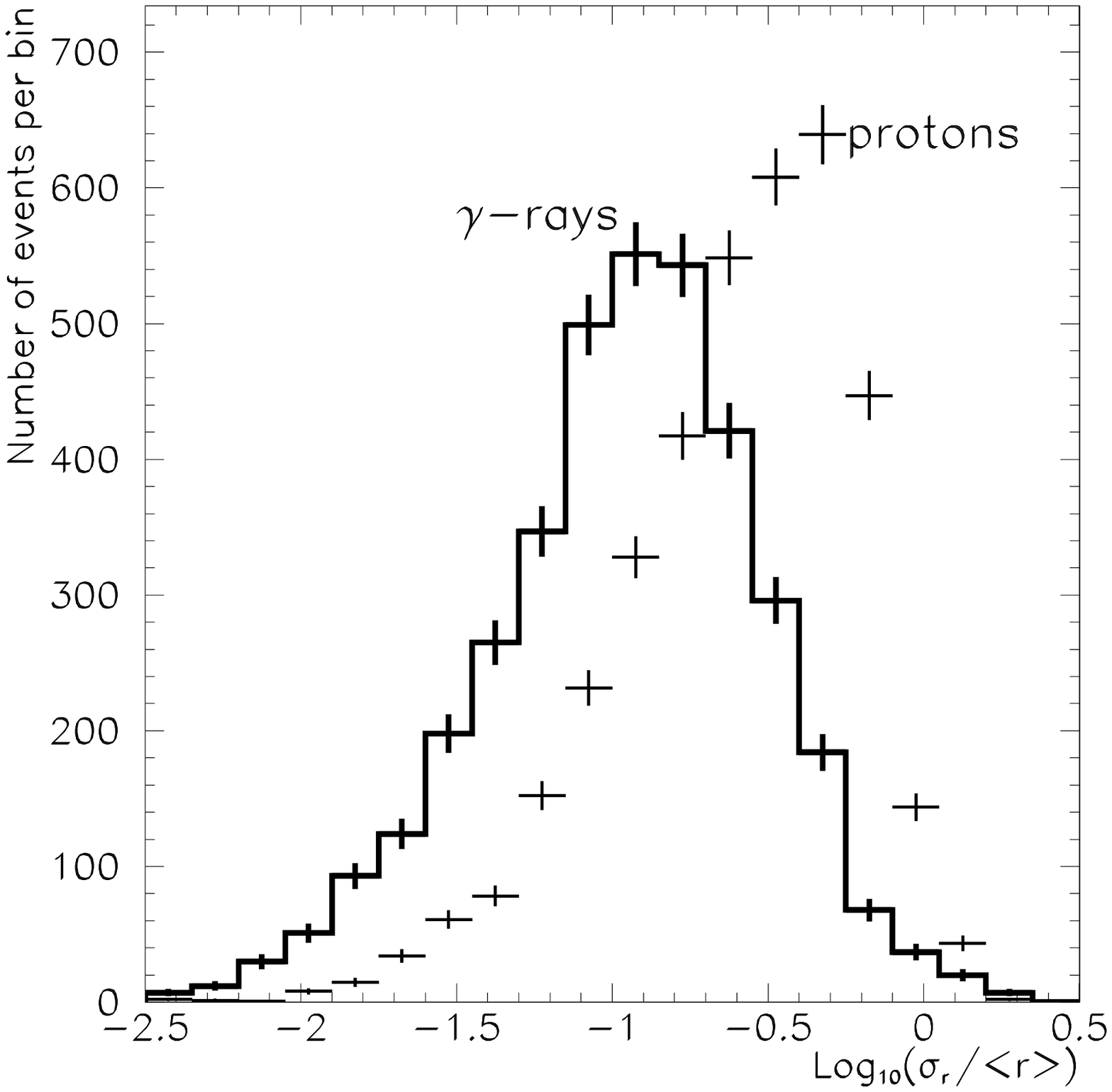,width=2.80in}
\epsfig{file=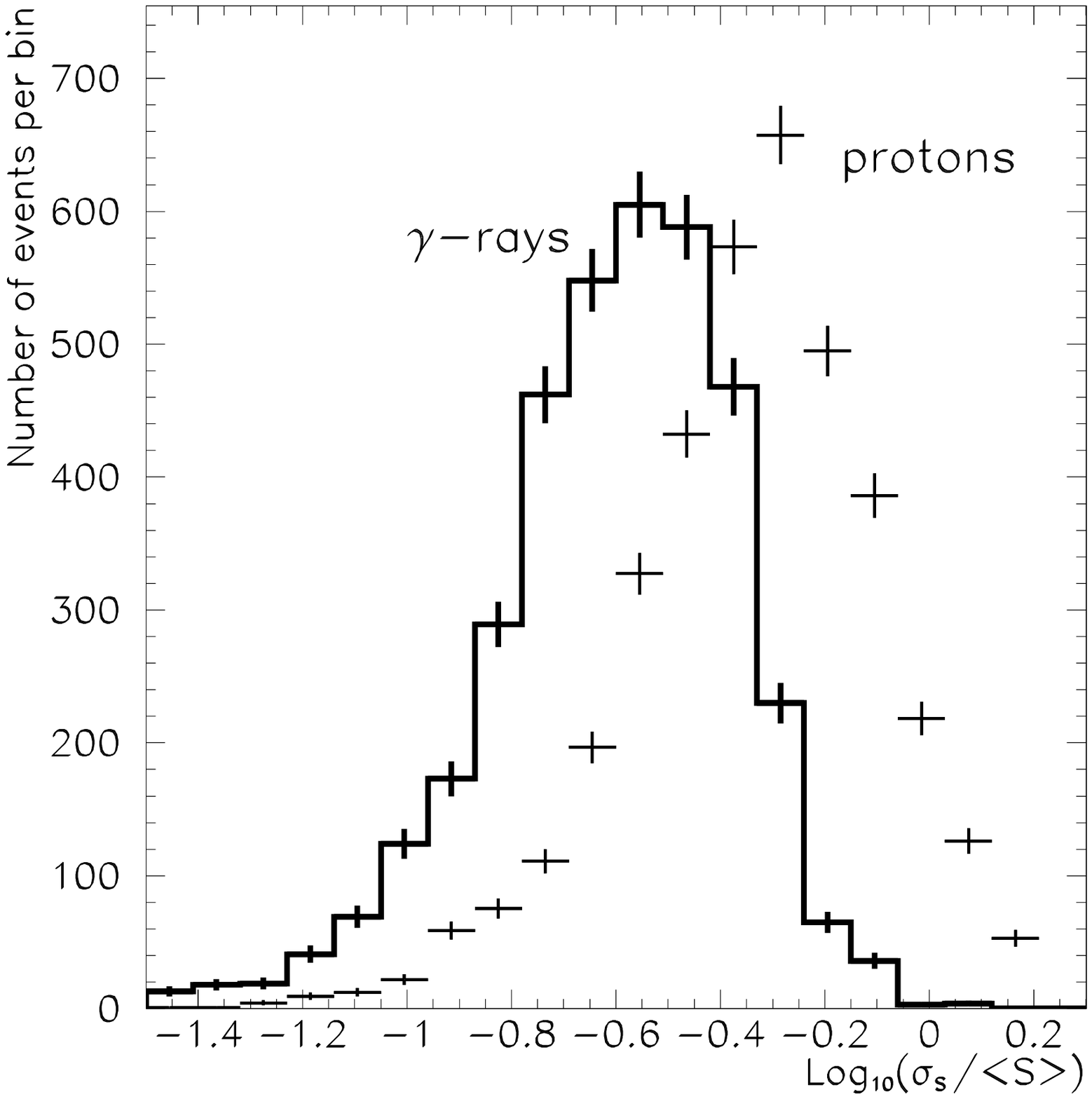,width=2.80in}
\vspace{10pt}
\caption{Distributions of the relative standard deviations of the impact 
parameter to source-image distance ratio (on the left) and of the image
luminosity for simulated 100-GeV $\gamma$-rays and 100-GeV to 10-TeV protons
recorded with a four-telescope array} 
\label{discri}
\end{figure}

In the case of point sources, the angular resolution provides an important 
factor of discrimination against cosmic-rays. Further discrimination is
possible using only the image centroid locations and the image brightness. 
A more detailed image analysis would be more efficient at rejecting 
cosmic rays, especially the high energy ones. Here we just concentrate on 
the exercise of only using the centroid position information and total 
image brightness.

The ratio $\rm r_i$, described in the previous section can
now be used as a discrimination tool against cosmic rays. 
As is often the case, different parts of a cosmic-ray 
shower are seen by different telescopes. Thus, the almost linear relationship 
between the $\rm (SC_i)$ and $\rm (PT_i)$ distances breaks down and the
$\rm r_i$ values will exhibit larger fluctuations between telescopes
for a given cosmic ray shower.
On the left panel of figure \ref{discri} we compare the distributions 
of the relative standard deviation, $\rm \sigma_r/ <r>$, of the $\rm
r_i$ values for simulated 100-GeV $\gamma$-rays and 
100-GeV to 10-TeV protons. A cut on the parameter $\rm \sigma_r/ <r>$
can preferentially select $\gamma$ rays. 

Additionally, we have not used the quantity of light $\rm s_i$ in each image. 
The amount of Cherenkov light received by each telescope from a $\gamma$-ray 
shower should follow the well-known Cherenkov light-pool density profile 
\cite{hillas96}. For vertical showers the profile is relatively flat
for impact parameters less than   
$\rm \sim 130~m$. On the contrary, cosmic rays produce 
Cherenkov light pools on the ground that are more patchy and irregular. On the 
right panel of Figure \ref{discri}, we compare the distribution of the
relative  
standard deviation of the quantity of light $\rm \sigma_s/<s>$. Here again a 
cut on this parameter permits some discrimination against cosmic rays. As an 
example, selecting 
$\rm Log_{10}(\sigma_r / <r>) < -0.72$ and 
$\rm Log_{10}(\sigma_s / <s>) < -0.43$ preserves $\rm 56\%$ of our simulated 
$\gamma$ rays while rejecting $\rm 86\%$ of the protons.

For an array of more than four telescopes ($\rm N>4$), the
reconstruction process provides a method for rejecting cosmic rays. With 
$\rm N>4$ the minimum of $\rm f(x_S,y_S)$ will generally not be zero. The 
value of the minimum will depend on how well the optimal source position and 
impact point account for the image centroid positions in all the N telescopes. 
Events for which the $\rm (SC_i)$ lines are not all precisely parallel
to the $\rm PT_i$  
lines, will have larger minimal values of $\rm f(x_S,y_S)$. As each  
telescope collects Cherenkov light from different  
parts of the cosmic-ray showers, cutting on the 
minimized value of $\rm f(x_S,y_S)$ provides one more discrimination
criterion. 

\section{Conclusion}

By using only the image centroid positions from imaging atmospheric
telescopes, the axis of $\gamma$-ray showers
can be reconstructed for arrays of at least four telescopes. For a
four-telescope array, we have verified that the analysis provides
reasonable precision. Applied to events close to detector
threshold, this method can provide a 
good starting point for more sophisticated
and much more computation-demanding methods involving, for example, 
detailed models 
of the Cherenkov images \cite{lebohec,deneroy03}. 

We have identified criteria useful for discriminating against cosmic rays 
while still using only the first-order image parameters, the image
centroids and brightnesses. These criteria could be combined
with more  
sophisticated image analysis, such as producing a global image width
from the light distribution across the line connecting the 
image centroid to the reconstructed source position.

Since the analysis presented here uses the minimal amount of
information from the recorded Chenenkov images, the results must
provide the lowest analysis energy threshold for a given
experiment. This analysis, in
comparison to analysis techniques using higher-order image parameters, 
could result in lower energy thresholds for modern air Cherenkov imaging
telescope arrays. This technique could also be used in the design of future
detectors and suggests that coarser pixels with perhaps smaller
light collectors than in HESS and VERITAS could be cost saving and
still produce accurate results in future very large arrays. 

\section{Acknowledgment}
We thank Dirk Putzfeld for his help and we acknowledge support by the 
Department of Energy High Energy Physics Division.

\end{document}